\documentclass[seceq]{ptptex}
\usepackage{graphicx,psfrag,amsmath,amssymb,amsfonts,bbm,bm}

\newcommand{\im}{i}

\title{
Spontaneous symmetry breaking: variations on a theme%
}

\author{
Giovanni \textsc{Jona-Lasinio}%
}

\inst{Dipartimento di Fisica, ``Sapienza'', Universit\`a di Roma, \\
Piazzale A. Moro 2, Roma 00185, Italy\\
Istituto Nazionale di Fisica Nucleare, Sezione di Roma 1, \\
Roma 00185, Italy}

\abst{Spontaneous symmetry breaking (SSB)  is a widespread phenomenon
in several areas of physics.
In this paper I wish to illustrate some situations where spontaneous 
symmetry breaking presents non obvious aspects. 
The first example is taken from molecular physics
and is related to the paradox of existence of chiral molecules. The second
case refers to a Dirac field  in presence of a magnetic field.
Gusynin, Miransky and Shovkovy have shown that  Nambu--Jona-Lasinio (NJL) 
models, where SSB
of chiral symmetry takes place for the nonlinear coupling over a certain
threshold, in presence of a magnetic field exhibit SSB for any value of 
the coupling. They called this phenomenon
{\sl magnetic catalysis}.  
I will discuss  this problem in $2+1$ dimensions
from an operatorial point of view and show that the basic 
phenomenon is a double {\sl pairing} induced by the magnetic field
in the vacuum for any value of the
mass. 
This pairing provides the {\sl environment} 
responsible for chiral symmetry breaking in NJL models with very weak 
nonlinearities.
The third case illustrates briefly SSB in stationary nonequilibrium states 
and its possible relevance in natural phenomena.}

\begin{document}

\maketitle

\section{Introduction}
The problems considered in this paper have a previous history in the 
literature. Here, after a brief review, we shall elaborate on their 
interpretation and significance.  

\medskip

The first problem refers to molecular physics, more precisely to the 
existence of chiral molecules, that is molecules that rotate the
polarization of light. Hund was the first to point out in 1927
that such molecules according to quantum mechanics should not exist
as the corresponding Hamiltonian is invariant under mirror reflection \cite{HU}.
This problem became known as the Hund's paradox and was discussed
in the physico-chemical literature for decades.

In the following we shall argue that SSB provides a solution
to the paradox once we take into account that  the experiments deal with a large
number of molecules and that their interactions are sufficient
to keep them in the chiral state. This idea was 
behind work done previously in collaborations with Pierre Claverie \cite{CJL},
Carlo Presilla and Cristina Toninelli \cite{JLPT}, but was not 
completely explicit
as the emphasis was on instabilities that set in at the semiclassical limit
and localization phenomena. We will mention also a complementary point
of view proposed by other authors. 

\medskip

The second case refers to a quantum Dirac  field in $2+1$ dimensions 
in presence of a magnetic field orthogonal to the plane where the
particles live. This system is invariant under chiral gauge transformations
in the limit of vanishing mass.  
Gusynin, Miransky and Shovkovy \cite{GMS} pointed out that 
in a NJL model \cite{NJL} in $2+1$ dimensions in presence of 
a magnetic 
field a  finite mass is produced, that is chiral SSB takes place,  
for any value of the nonlinear coupling. 
In other words there is no
threshold as it happens in absence of magnetic field. They showed
that  the same happens in $3+1$ dimensions \cite{GMS1}. 
They called this phenomenon \emph{magnetic catalysis} and explained 
it in terms of a 
\emph {dimensional reduction} which modifies the
infrared behaviour of the system. 

In this paper, I will resume an explicit expression of the vacuum of a 
Dirac  field 
in $2+1$ dimensions in presence of a magnetic field obtained in collaboration
with Francesca Marchetti long ago \cite{JLM} and will emphasize that the
magnetic catalysis is ultimately due to the fact that the magnetic 
field induces for any value of the mass of the Dirac field, 
including zero mass, a  pairing structure in the vacuum 
typical of NJL type models with SSB. 
Therefore once the magnetic field is switched on the
pairing is given for free and SSB in NJL models may take  place with an 
arbitrarily small nonlinear interaction. 

The pairing structure 
of the vacuum induced by the magnetic field is similar in $2+1$ and $3+1$ 
dimensions. This, I believe, is at the origin of the similar behavior 
of NJL models in presence of a magnetic field in these dimensions.
 
A special feature of magnetic pairing, as we shall see, is that there
are two levels at which pairing is established. This is apparently 
a new structure which deserves further study.

\medskip

The third case is a new opportunity provided by a discovery of SSB 
in very simple
models of nonequilibrium statistical mechanics \cite{NEQ,PEM,NEQ2}, 
studied also for their interest in connection with traffic problems.
The models studied deal with two types of particles, e.g. positive and 
negative, moving on a one-dimensional lattice. In equilibrium the states of 
the system are symmetric under the exchange of the two types of particles. 
Out of equilibrium, e.g. by putting different injection rates at the two 
boundaries of the lattice, the system is invariant under CP, the combined 
action of charge conjugation 
and parity. However, both symmetric and symmetry breaking steady states are 
possible depending on the values of these rates. In the symmetry breaking 
states different densities and currents are present for the two types of 
particles.
  
Here we shall point out the possible relevance of this phenomenon 
in relation with well known unsolved problems in molecular biology
and cosmology.

\section{On the existence of chiral molecules}

The behavior of gases of pyramidal molecules, 
i.e. molecules of the kind $XY_3$ like ammonia $NH_3$, phosphine $PH_3$,
arsine $AsH_3$,
has been the object of investigations since the early developments
of quantum mechanics \cite{HU}.
Suppose that we replace two of the hydrogens with different atoms like
deuterium and tritium: we obtain a molecule of the form $XYWZ$.
This is called an enantiomer, that is a molecule whose mirror image
cannot be superimposed to the original one. These molecules
are optically active in states where the atom $X$ is localized on one side
of the $YWZ$ plane. However according
to quantum mechanics such localised states should not exist as
stable stationary states. In fact the two possible positions of the
$X$ atom, usually separated by a potential barrier, are accessible to $X$ via 
tunneling giving rise to wave functions delocalized over the two
minima of the potential and of definite parity. In particular
the ground state is expected to be even under parity. Tunneling 
induces a doublet structure of the energy
levels. The level splitting is in the microwaves for ammonia but
it is not detectable in phosphine and arsine. In the latter case 
the estimated splitting is very  small and, as far as I know,
is not experimentally
accessible. However empirical evidence shows that  there are optical isomers
of arsine in which the $As$ atom  must be localized in one of the minima. 

Different approaches have been proposed to explain the localization.
A common feature is the basic role attributed to the environment of
a molecule which is never isolated. The environment can be made
by molecules identical to the one under consideration, or different.
The approaches differ as one can tackle the problem either from
a dynamical or a static point of view. The dynamical view has been
advocated for example in \citen{MS,HS} and more recently in \citen{TH}.    
The static or equilibrium approach has been developed mainly
in the papers \citen{CJL,JLPT}. 
The comparison between the two points of view is not so simple  
and we shall comment on this question later.  

The situation considered in the static theory is that
of a gas of weakly interacting identical pyramidal molecules at room 
temperature. 
For molecules with sufficiently large dipole moments the so-called
Keesom forces are supposed to dominate the intermolecular interactions.
These are forces between two dipoles not fixed in orientation
averaged with the Boltzmann factor at temperature $T$.
A quantitative discussion of the collective effects induced by coupling 
a molecule to the environment constituted by the other 
molecules of the gas was made in \citen{CJL}.
In this work it was shown that, due to the instability
of tunneling under weak perturbations,
the order of magnitude of the molecular dipole-dipole interaction 
may account for localized ground states.

This suggested that a transition to localized states should happen 
when the interaction among the molecules is increased.
Evidence for such a transition was provided by
measurements of the dependence of the doublet frequency under
increasing pressure which vanishes for a critical pressure $P_{cr}$
different for  $NH_3$ and $ND_3$. The measurements were taken
at the end of the forties and beginning of the fifties \cite{BL,BM}
but no quantitative theoretical explanation was given for 
fifty years. Then a simple model was proposed in \citen{JLPT}
wich gave a satisfactory account of the empirical results
and a transition emerged at the critical pressures.
Although this was not particularly emphasized in that paper,
SSB is a natural concept to describe the transition even though
some peculiarities have to be pointed out.
A remarkable feature of the model is that there are no
free parameters. Before presenting the model I wish to
describe the tunneling instability underlying the localization
phenomenon.
 
\subsection{\bf {Tunneling instability in the semiclassical limit}}
In this  subsection we describe the instability of
tunneling  under weak perturbations leading to a localization
of a molecule under semiclassical conditions \cite{JMS,JMS1}. 
We argue that this phenomenon reminds of a \emph{phase transition}.  

Let us consider a symmetric double well potential $V_0(x)$, e.g. 
$V_0(x)=V_0 \,(x^2-1)^2$ where $V_0 > 0$ is the height of the barrier
separating two two minima at $x=\pm 1$. Let
a perturbing potential $V_1(x)$ be localized inside one of the wells
but possibly away from the minimum. More precisely 
\begin{eqnarray}
V_1(x) \neq 0, \quad x \in (a_1,a_2) \subset (0,x^*), \quad
V_1(x)=0 \quad \text{otherwise} \;.
\end{eqnarray}
The interval $(0,x^*)$ includes the minimum at $x=1$
and $(a_1,a_2)$ is a small segment compared to $(0,x^*)$ so that the
perturbation modifies only locally the double well.
Then, essentially independently of the strength of the perturbation,
the following estimate holds for sufficiently small $\hbar/m$, where
$m$ is the mass of the tunneling particle,
\begin{eqnarray}
{\frac {\psi_0 (1)} {\psi_0 (-1)}} \approx -{\frac {\psi_1 (-1)} {\psi_1 (1)}}
\approx  \exp{[- \frac 1 \hbar 
\int_{-a_2}^{a_2} (2mV_0(x))^{1/2} dx]} \;.
\end{eqnarray} 
Here $\psi_0 (x)$ and  $\psi_1 (x)$ denote respectively, the ground state
and the first excited state of the \emph{perturbed} problem.

The \emph{independence} of this estimate on the intensity of the perturbation
holds provided $V_1 \gg A\exp[-\frac {C(a_2)} {\hbar}]$ where
$C(a_2)=2\int_0^{a_2} (2mV_0(x))^{1/2} dx$ and $A$ is a pre-factor having
the dimension of an energy. 

The meaning of this result is that we expect the tunneling atom in a 
non-isolated pyramidal molecule to be generically localized 
under semiclassical
conditions, that is $(mV_0)^{1/2} \Delta x/\hbar \gg 1$, with $V_0$
the height of the barrier and $\Delta x$ his width. It is well known that
changing the curvature of one of the two symmetric minima produces localisation
in the well with the smallest curvature. The above result shows that local
much weaker perturbations have the same effect. The sign of $V_1$
determines the well where localisation occurs.

From the standpoint of a functional integral description the particle in
a double well is like a one-dimensional system of continuous spins 
and it cannot exhibit a phase transition at finite temperature.
Note that in our case $\hbar$ has  the same role as the temperature in  
statistical mechanics.
The phenomenon described however is similar to what happens in  the Ising model
in $2d$ below the critical point where it is extremely sensitive
to boundary conditions.
Boundary conditions, like our local perturbations, act on a space scale 
small compared to the bulk but are sufficient to 
drive the system in a state of definite magnetization \cite{GAL}.

\subsection{\bf {Localisation and disappearance of the inversion line}}
If we deal with a set of molecules (e.g. in the gaseous state), once
localization takes place for a molecule there appears a cooperative effect
which tends to stabilize this localization. The mechanism is called the
reaction field mechanism. Let $\mu$ be the dipole moment of the localized 
molecule; this moment
polarizes the environment which in turn creates the {\sl reaction field}
$\cal E$ which is collinear with $\mu$ and the interaction energy,
that plays the role of the perturbation $V_1$ of the previous subsection, 
$V_I = - {\frac 12}\mu \cdot \cal  E$ is negative.
If $|V_I| >> \Delta E$, where $\Delta E$ is the doublet splitting 
due to tunneling in the isolated symmetric state,  the molecules 
of the gas are localized. As a consequence the doublet should
disappear when $|V_I|$ increases for example by increasing the pressure. 

\medskip

We summarize the results of \citen{JLPT}.
In the physical situation we are considering the one-dimensional 
inversion motion 
of the nucleus $X$ across the plane containing the three nuclei $Y$
can be separated from the rotational degrees
of freedom. Centrifugal forces due to rotation simply change the
effective potential in which the molecule vibrates.
The form of the effective potential for this motion is 
a double well which is symmetric with respect to the inversion plane 
\cite{W,T}. 

For the pyramidal molecules under consideration,
the thermal energy $k_BT$ at room temperature is much smaller 
than the distance between the first and the second doublet so that  
the problem can be reduced to the study of a two-level system 
corresponding to the symmetric and anti-symmetric states of the 
first doublet. 

In \citen{JLPT} we mimicked the inversion degree of freedom of an isolated 
molecule with 
the Hamiltonian
\begin{equation}
h_0=-\frac{\Delta E}{2}\sigma^x \;,
\end{equation}
where $\sigma^x$ is the Pauli matrix in the standard representation
with delocalized tunneling eigenstates
\begin{equation}
|1\rangle = \frac{1}{\sqrt 2}
\left( \begin{array}{c} 1 \\ 1 \end {array} \right) 
\qquad
|2\rangle = \frac{1}{\sqrt 2} 
\left( \begin{array}{c} \phantom{-}1 \\ -1 \end {array} \right) \ .
\label{12}
\end{equation}
Since the rotational degrees of freedom of the single pyramidal molecule 
are faster than the inversion ones, 
on the time scales of the inversion dynamics the molecules 
feel an effective attraction arising from the angle averaging of the 
dipole-dipole interaction at the temperature of the experiment. 

\medskip

In the representation chosen for the Pauli matrices, 
the localizing effect of the dipole-dipole interaction between two
molecules $i$ and $j$ can be represented by an interaction term 
of the form $\sigma^z_i \sigma^z_j$, 
where $\sigma^z$ has localized eigenstates
\begin{equation}
|L\rangle = 
\left( \begin{array}{c} 1 \\ 0 \end {array} \right) 
\qquad
|R\rangle = 
\left( \begin{array}{c} 0 \\ 1 \end {array} \right) \ .
\label{LR}
\end{equation} 
In a mean-field approximation we obtain the total Hamiltonian
\begin{equation}
h(\lambda)=-\frac{\Delta E}{2}\sigma^x-
G\sigma^z\langle\lambda|{\sigma^z}|\lambda\rangle \;,
\label{acca}
\end{equation}
where $|\lambda\rangle$ is the single-molecule state to be determined
self-consistently.

\medskip

The parameter $G$ represents the average dipole interaction energy of a 
single molecule with the rest of the gas.
This must be identified with a sum over all possible molecular 
distances and all possible dipole orientations calculated with the 
Boltzmann factor at temperature $T$.
If $\varrho$ is the density of the gas, we have
\begin{equation}
G=\int_d^\infty  
\frac{\mu^4}{3(4\pi \varepsilon_0\varepsilon_r)^2 k_BT r^6}
~\varrho~4\pi r^2 \mathrm{d}r \;,
\label{GG}
\end{equation}
where $\varepsilon_0$ is the vacuum dielectric constant, $\varepsilon_r$ 
the relative dielectric constant, $k_B$ the Boltzmann constant and  
$d$ the molecular collision diameter. 
The fraction in the integrand represents the Keesom energy between 
two classical dipoles of moment $\mu$ at distance $r$.
Equation (\ref{GG}) is valid in the range of temperatures
appropriate for room temperature experiments.

\medskip

By defining the critical value $G_\mathrm{cr}=\Delta E/2$, 
we distinguish the following two cases.
For $G<G_\mathrm{cr}$, 
the ground state of the system is approximated by a 
product of delocalized symmetric single-molecule states 
corresponding to the ground state of an isolated molecule.
For $G \geq G_\mathrm{cr}$, we have two different product states 
which approximate the ground state of the system. 
The corresponding single-molecule states transform 
one into the other under the action of the inversion operator $\sigma^x$, 
and, for $G \gg G_\mathrm{cr}$, 
they become localized
\begin{equation}
\lim_{\Delta E/G \rightarrow 0}|\lambda_0^L\rangle=|L\rangle
\qquad
\lim_{\Delta E/G \rightarrow 0}|\lambda_0^R\rangle=|R\rangle \ .
\label{loc}
\end{equation}

The above results imply a bifurcation of the ground state 
at a critical interaction $G=G_\mathrm{cr}=\Delta E/2$. 
Using the equation of state for an ideal gas $\varrho=P/k_BT$, 
this bifurcation can be related to the
increasing of the gas pressure above a critical value 
\begin{equation}
P_\mathrm{cr}=\frac{9}{8 \pi} P_0 \left(\frac{T}{T_0}\right)^2 \;,
\label{pcr}
\end{equation}
where $P_0=\Delta E/d^3$ and 
$T_0=\mu^2/(4\pi \varepsilon_0\varepsilon_r d^3 k_B)$ \ .
In fact by expressing $\varrho$ in \eqref{GG} with the equation of state
and setting $G=G_\mathrm{cr}$ we obtain \eqref{pcr} and $P > P_\mathrm{cr}$
means $G > G_\mathrm{cr}$.

\medskip

When the gas is exposed to an electro-magnetic radiation of
angular frequency $\omega_0$, using the linear response theory
in a dipole coupling approximation, in \citen{JLPT} we obtained the theoretical 
expression for the inversion line frequency as a function of
pressure
\begin{equation}
\bar\nu=\frac{\Delta E}{h}\sqrt{1-\frac{P}{P_\mathrm{cr}}} \;,
\label{cicacica}
\end{equation}
where $P_\mathrm{cr}$ is given by (\ref{pcr}).
Note that this expression does not contain free parameters.
 
\medskip

Equation (\ref{cicacica}) predicts that, 
up to a  pressure rescaling, the same behavior of $\bar\nu(P)$ 
is obtained for different pyramidal molecules
\begin{equation}
\frac{\bar\nu_{XY_3}(P)}{\bar\nu_{XY_3}(0)}=
\frac{\bar\nu_{X'Y'_3}(\gamma P)}{\bar\nu_{X'Y'_3}(0)} \; ,
\label{pgamma}
\end{equation}
where
$\gamma = \left. P_\mathrm{cr}\right._{X'Y'_3}/
\left.P_\mathrm{cr}\right._{XY_3}$.

We compared our theoretical analysis of the inversion 
line with the spectroscopic data available for 
ammonia and deuterated ammonia \cite{BL,BM}.
In these experiments the absorption coefficient of a cell
containing $NH_3$ or $ND_3$ gas at room temperature   
was measured at different pressures.
The frequency $\bar{\nu}$ of the inversion line decreases by increasing 
$P$ and vanishes for pressures greater than a critical value.
In the case of $ND_3$ and $NH_3$, at the same temperature $T$ we have 
$\gamma= \Delta E_{NH_3}/\Delta E_{ND_3} \simeq 15.28$.
This factor has been used to fix the scales of the figure.
We see that in this way the $NH_3$ and $ND_3$ data fall on the same curve.

\begin{figure}   
  \begin{center}
  \includegraphics[width=12cm,clip]{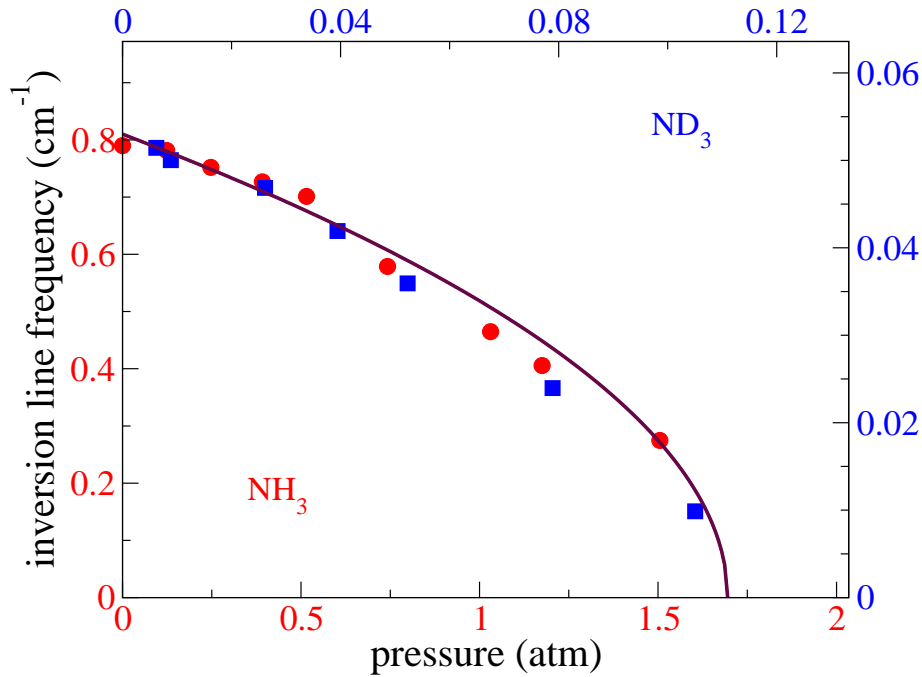}
  \end{center}
  \caption{Measured inversion-line frequency $\bar\nu$ 
    as a function of the gas pressure $P$
    for $NH_3$ (dots, left and bottom scales, data from
    \protect{\citen{BL}})
    and $ND_3$ (squares, right and top scales, data from 
    \protect{\citen{BM}}). 
    The solid line is the theoretical formula (\protect{\ref{cicacica}}) 
    with $P_\mathrm{cr} = 1.695$ atm for $NH_3$ 
    and $P_\mathrm{cr} = 0.111$ atm for $ND_3$
    calculated according to (\protect{\ref{pcr}}).}
\end{figure}
 
\medskip

At pressures greater than the critical one we have the following 
situation.
In the limit of an infinite number of molecules, 
the Hilbert space separates into two sectors generated by the ground 
state vectors given in mean field approximation by
\begin{eqnarray}
|\psi_0^L\rangle&=&|\lambda_0^L\rangle\ldots|\lambda_0^L\rangle\\
|\psi_0^R\rangle&=&|\lambda_0^R\rangle\ldots|\lambda_0^R\rangle \ .
\end{eqnarray}
These sectors, which we call $\cal{H}_L$ and $\cal{H}_R$,  
cannot be connected by any operator involving a finite number of 
degrees of freedom (local operator). 
This means that a superselection rule operates between them.

\subsection{\bf Chiral molecules as a case of SSB}
The mean field model outlined in the previous subsection has the
essential features of SSB however with some peculiar aspects.
Our model describes, according e.g. to the definition of \citen{SS},
a  quantum phase transition 
involving only the inversion degree of freedom of the molecules. 
We have assumed in fact a separation
of the inversion dynamics from the rotational and translational
degrees of freedom. 
In this respect there may be a difference
between a molecule of the form $XY_3$ and an optically active molecule
$XYZW$ which is less symmetric and a coupling with rotational degrees of
freedom is to be expected. However according to \citen{T}, empirically
the decoupling appears to work reasonably well also in this case. 
In view of the success of our model in reproducing the experimental
results in spite of the simplification of the actual system, an effort
must be made to derive it from the complete dynamics involving all
the degrees of fredom of the molecules. Work is in progress on this
problem. Furthermore, considering that the measurements were taken
long ago, new experiments seem desirable.

A similar treatment can be applied to  a gas of 
pyramidal molecules of very low density in an environment
provided by a gas of higher density of molecules of a different kind.
In this case one may neglect the interaction among
the pyramidal molecules and reduce the problem to a single
molecule interacting with the environment. If this is made
of units  having a dipole moment the situation is not
substantially different from the one discussed above. If
the molecules of the environment are not endowed with a
dipole moment they can acquire it through deformation under
the action of the  field of the pyramidal molecules. This is again
the reaction field mechanism.  Of course
the localizing effect is much weaker and, for example, in a
gas of $He$ the critical pressure for $NH_3$ is of the order of 
$7.3 \cdot 10^3 \: atm$
which is still experimentally accessible in the laboratory. 

In a recent work by
Trost and Hornberger \cite{TH} the localization problem is  analysed 
in  detail from 
a dynamical point of view. The basic idea of the mechanism is that 
through repeated
scatterings in a host gas producing decoherence, a molecule is blocked
in a localized state. Dissipation is involved in the process.   
It is not easy to compare the dynamical approach with our theory.
Let us consider the previous case of a pyramidal molecule
in a non-chiral gas. From our point of view the localized
state of the molecule, while not an eigenstate of the isolated molecule,
corresponds to a stationary state of the system formed by the molecule 
and the  gas.
Other situations where this point of view turns out to be useful
were discussed in \citen{JLC}.   
If I understand correctly 
the dynamical point of view, the molecule remains in a non stationary
state for ever due the environment but the stationary states of the 
composite system
are never considered. In \citen{MS} Simonius claimed that the
dynamical blocking of states {\sl ``differs from, and is much more powerful
than, the one [SSB] usually discussed in the literature based on nonsymmetric
solutions to symmetric equations''}.  
In my opinion, there is not yet a satisfactory  understanding of when
the decoherence picture applies. For example  in conditions
of temperature and pressure in which the inversion line due to tunneling
is observed the dynamical mechanism must be ineffective. However one
can conceive situations in which the
two pictures are complementary descriptions of the same reality. 
     
We emphasize that so far the SSB interpretation of molecular chirality,
besides being conceptually simple, provides for the first time
a theory having a  qualitative and quantitative support 
from the empirical data. These matters will be discussed in detail in a 
forthcoming paper  \citen{JLP}. For a general assessment of the 
problem of chiral molecules see also \citen{ASW}.
 
\section{\bf {Magnetic pairing}}
The presence of a magnetic field can generate new
interesting phenomena in three-dimensional gauge theories as was
recognized long ago by Hosotani \cite{HO}. At about the same time
Gusynin, Miransky and Shovkovy discussed chiral symmetry breaking
in NJL models with magnetic field as mentioned before.
The phenomenon of chiral symmetry breaking in presence of an external  
homogeneous magnetic field of an internal $U(2)$ symmetry for the  
Dirac field in $2+1$ dimensions has recently attracted new  
attention. The reason why people are interested in this phenomenon  
is its possible relevance in the physics of two-dimensional
systems like graphene, see e.g. \citen{KY,RMP} and references therein.

In the following we discuss the construction of the Hilbert space
for a Dirac field in $2+1$ dimensions under the action of an external
magnetic field for any value of the mass. This is done in two steps
which exhibit  two levels of pairing. There is a first level in
which pairs of operators of opposite momentum in the fourier
development of the free field appear, giving rise
to a new Fock space. In this space we introduce  
creation and destruction operators associated to the wave functions
characterizing the Landau states which depend on a discrete index $n$. 
A subsequent pairing of such operators 
leads to a second Fock space which is the Hilbert space
of the full relativistic problem. It is a rather peculiar structure 
which, as far as I know, was not encountered before.

\medskip

\subsection{\bf {Preliminaries}}
In the following we use units with $\hbar=c=1$ where $c$ is the velocity 
of light, as their values do not play a role in our discussion.
Consider the Lagrangian $\mathcal{L}$: 
\begin{equation} 
\begin{split} 
{\mathcal L}     &= {\bar \psi}^B (x) \left[ \im \gamma^\mu 
{\mathcal D}_\mu - m  \right] \psi^B (x) \\ 
{\mathcal D}_\mu &= \partial_\mu - \im e A_\mu \;, 
\end{split}
\label{elle1}
\end{equation}
where ${\bar \psi}^B (x)$ is the Dirac quantized field  in presence of 
the vector potential $A_\mu$
and $e$ is the modulus of the electron charge. Since the 
magnetic field is constant and homogeneous we can choose 
the Landau gauge:   
\begin{equation} 
\label{gaug1} A_\mu = -\delta_{\mu 1} Bx_2 \ .
\end{equation}
Notice that introducing the magnetic length $l=1/(eB)^{1/2}$ and using it
as a unit for the space time variables, the magnetic field
can be rescaled to the value $1$.
However in the following we shall keep
the magnetic field explicit.

The problem of a free Dirac field, minimally coupled to a 
homogeneous magnetic field, can be exactly solved and in the 
Landau gauge the expression of the Dirac field in the so called chiral 
version \cite{GMS,JLM} is ($eB>0$): 
\begin{gather}
\label{psim1}
\psi^B (x) = \begin{pmatrix} \psi_1^B (x) \\ \psi_2^B (x) 
\end{pmatrix} \\
\label{fourier} 
\psi_1^B (x) = \sum_{n=0}^{\infty} \sum_{p_1} \{u_{np_1}(x) 
a_{np_1} + v_{n-p_1}(x) b_{np_1}^\dag \} \\ 
\psi_2^B (x) = \sum_{n=0}^{\infty} \sum_{p_1} \{ u_{n p_1}^{(2)} 
(x) c_{np_1} + v_{n -p_1}^{(2)} (x) d_{np_1}^\dag \} 
\label{psim2}
\end{gather}
\begin{equation}
\begin{split}
u_{np_1}(x) &= \frac{1}{\sqrt{lL_1}}e^{-\im E_nt}e^{\im p_1 x_1} 
\begin{pmatrix} A_n w_n(\xi_{x_2}^{p_1}) \\ -\im B_n w_{n-
1}(\xi_{x_2}^{p_1}) \end{pmatrix} \\ 
v_{np_1}(x) &= \frac{1}{\sqrt{lL_1}}e^{+\im E_nt}e^{\im p_1x_1} 
\begin{pmatrix} B_n w_n(\xi_{x_2}^{p_1}) \\ +\im A_n w_{n-
1}(\xi_{x_2}^{p_1}) \end{pmatrix} 
\end{split}
\label{enepm}
\end{equation}
\begin{equation}
\label{uv2}
u_{np_1}^{(2)} (x) = (-1)^n v_{n-p_1} (-x) \quad v_{np_1}^{(2)} 
(x) =(-1)^n u_{n-p_1} (-x) 
\end{equation}
\begin{equation}
A_n = \sqrt{\frac{E_n + m}{2 E_n}} \qquad B_n = \sqrt{\frac{E_n - 
m}{2 E_n}} 
\label{defin}
\end{equation}
\begin{equation}
E_n = \sqrt{m^2 + 2neB} \qquad \xi_{x_2}^{p_1} = \frac{x_2}{l} + 
lp_1 = \sqrt{eB} x_2 + \frac{p_1}{\sqrt{eB}} 
\end{equation}
\begin{equation}
w_n (\xi) = c_n e^{-\xi_2 / 2} H_n (\xi) = \left( 2^n n!\sqrt{\pi} \right)^{-1/2} e^{-\xi^2 / 2} H_n (\xi) \ . 
\end{equation}
The operators $a_{n p_1} , ... , d_{np_1}$ satisfy  
canonical anticommutation relations and $H_n (\xi)$ are the 
Hermite polynomials. The case $eB<0$ can be obtained by applying 
the charge conjugation operator. For a discussion of the symmetry 
properties of the problem we refer to  \citen{JLM,GMS}. Here we only
mention that apart from the mass term the Lagrangian is invariant
under chiral transformations and 
in the limit $m \rightarrow 0$ in $2+1$ dimensions there is
spontaneous
symmetry breaking revealed by the calculation of the order parameter
\begin{equation} 
\lim_{m \to 0} \langle B| \bar{\psi}^B (x) \psi^B (x) |B \rangle = 
- {\frac{|eB|}{2\pi}}. 
\label{prima}
\end{equation} 

\medskip

The problem solved in \citen{JLM} was the calculation of the formal 
relationship between
the Dirac field in presence of magnetic field and the free Dirac field,
\begin{equation}
  \psi_1 (x) = \sum_{\bm{p}} \sqrt{\frac{m}{L_1L_2 E_{\bm{p}}}} 
  \left\{ u(\bm{p}) e^{-\im p \cdot x}a_{\bm{p}} + v(\bm{p}) 
    e^{\im p \cdot x} b_{\bm{p}}^\dag \right\} 
\label{svicm}
\end{equation}
 \begin{equation}
   u(\bm{p}) = \sqrt{\frac{E_{\bm{p}} + m}{2m}}
   \begin{pmatrix} 1 \\ \frac{p_2-\im p_1}{E_{\bm{p}} + m} 
   \end{pmatrix}  
   \quad v (\bm{p}) = \sqrt{\frac{E_{\bm{p}} + m}{2m}}
   \begin{pmatrix} 
     \frac{p_2+\im p_1}{E_{\bm{p}} + m} \\ 1 
   \end{pmatrix} \;, 
   \label{ugiug}
 \end{equation} 
where $E_{\bm{p}} = (m^2 + |\bm{p}|^2)^{1/2}$. Similarly for $\psi_2$.

Following \citen{NJL} we obtained the desired relationship
by imposing the same initial condition on the
Dirac equations describing the free field 
and the one in presence of the external magnetic field, 
\begin{equation}
\psi_i (0,{x}) = \psi_i^B (0,{x}) \quad i=1,2 \ .
\label{conin}
\end{equation}

\subsection{\bf {Constructing the Hilbert space}}
The following presentation reverses to some extent the order of the 
calculation in \citen{JLM} but this helps in bringing out the main points.
From \citen{JLM} it is clear that there is a
natural ambient  space for the construction of the
vacuum of the Dirac field in $2+1$ dimensions with constant
magnetic field.  
Let us define 
\begin{eqnarray}
A_p= \sqrt{\frac {E_p + m}{2Ep}}(a_p - C_p b^{\dag}_{-p})\\
B_p=\sqrt{\frac {E_p + m}{2Ep}}(C_pa^{\dag}_p +  b_{-p}) \;,
\end{eqnarray}
with
\begin{equation}
C_{\bm{p}} = \frac{p_2 + \im p_1}{E_{\bm{p}} + m} \qquad.
\end{equation}
The operators $A_p, A^\dag_p$ and $B_p, B^\dag_p$ satisfy the usual canonical
anticommutation relations (CAR). The
corresponding vacuum is
\begin{eqnarray}
|\hat 0\rangle=\prod_{\bm{h}} \frac{E_{\bm{h}} + m}{2 E_{\bm{h}}} 
\left( 1 + C_{\bm{h}} a_{\bm{h}}^\dag b_{-\bm{h}}^\dag 
\right)|0 \rangle \ .
\end{eqnarray}
By applying
the operators $A_p^\dag,B_p^\dag$ to $|\hat 0 \rangle$ we generate a 
new Fock space where a first pairing appears. Notice that
pairs carry a phase which depends on the momentum.

We next introduce the creation and destruction operators
\begin{equation} 
\begin{split} 
\widehat{a}_{np_1}=A(f_{np_1})  &=\sum_{p_2} f_{np_1}(p_2) A_{p} \\ 
 \widehat{b}_{n -p_1}=B(-if_{n-1\;-p_1}) &= \sum_{p_2} 
-if_{n-1\;-p_1}(p_2)B_{p} \;,  
\end{split}
\label{defos}
\end{equation}
where 
\begin{equation} 
f_{np_1}(p_2) = i^n\sqrt{2\pi l} e^{-\im l^2 p_1 p_2} w_n(lp_2) \;, 
\label{defcd}
\end{equation}
are the fourier transform in one variable of the wave functions 
appearing in \eqref{enepm}.

The main result of \citen{JLM} is the identification of 
the operators $a_{np_1},b_{np_1}$ appearing
in  \eqref{fourier} with those defined by the Bogolyubov transformation
\begin{equation}
\left\{
  \begin{aligned}
    a_{np_1} &= A_n \widehat{a}_{np_1} - B_n \widehat{b}_{n -p_1}^\dag 
    \\ 
    b_{n -p_1} &= A_n \widehat{b}_{n -p_1} + B_n \widehat{a}_{n 
      p_1}^\dag \ . 
  \end{aligned}
\right.
\label{refa1}
\end{equation}
A similar analysis can be done for $\psi_2^B$ introducing operators
$\widehat{c}_{np_1}$ and $\widehat{d}_{np_1}$ and using the equations
\eqref{psim2},  \eqref{uv2}.

The previous discussion then 
leads to the following expression for the vacuum of the Dirac field 
in presence of a constant magnetic field 
\begin{equation} 
|B \rangle = \prod_{n \ge 1} \prod_{p_1} \left( A_n + B_n 
{\widehat{a}_{np_1}}^\dag {\widehat{b}_{n-p_1}}^\dag \right) 
\left( A_n - B_n {\widehat{c}_{np_1}}^\dag {\widehat{d}_{n-
p_1}}^\dag \right)|\hat 0\rangle \;,
\end{equation}
where, 
\begin{equation}
|\hat 0\rangle=\prod_{\bm{h}} \frac{E_{\bm{h}} + m}{2 E_{\bm{h}}} 
\left( 1 + C_{\bm{h}} a_{\bm{h}}^\dag b_{-\bm{h}}^\dag 
\right) \left( 1 + C_{\bm{h}}^\ast c_{\bm{h}}^\dag d_{-
\bm{h}}^\dag \right) |0 \rangle \ . 
\end{equation}
We emphasize once more that this expression holds for any value of $m$.
By applying the conjugate of the operators defined in \eqref{refa1},
and the corresponding ones associated to $\psi_2$, to the vacuum  
$|B\rangle$ we generate the full Hilbert space of the Dirac
field in a magnetic field.

A simple calculation shows that 
$\langle0|\hat 0\rangle=0$ and $\langle\hat 0|B\rangle=0$.
Therefore three Fock spaces orthogonal to each other are
involved in the diagonalization of
the Hamiltonian of a Dirac field in presence of a
constant magnetic field in $2+1$ dimensions. 
  
\medskip

In order to clarify the meaning of the hatted operators, following \citen{JLM},
let us write the Hamiltonian in presence of magnetic field in terms of 
these operators. Using \eqref{refa1} we find 
\begin{equation}
\begin{split}
\displaystyle \mathrm{H}^B &= \widehat{\mathrm{H}} - 
\sum_{n=1}^\infty \sum_{p_1} \sqrt{2neB} \left( 
{\widehat{a}_{np_1}}^\dag {\widehat{b}_{n-p_1}}^\dag + 
\widehat{b}_{n-p_1} \widehat{a}_{np_1} \right) \\ 
\displaystyle \widehat{\mathrm{H}} &\equiv m \sum_{n=0}^\infty 
\sum_{p_1} \left( {\widehat{a}_{np_1}}^\dag \widehat{a}_{np_1} + 
{\widehat{b}_{np_1}}^\dag \widehat{b}_{np_1} \right) \ . 
\end{split}
\label{mamma}
\end{equation}

The expression of the Hamiltonian $\widehat{\mathrm{H}}$ shows that 
$\widehat{a}_{np_1}$ and $\widehat{b}_{np_1}$ describe particles 
with degenerate energy $m$, that is the energy of the 
lowest Landau level (LLL). The 
magnetic field induces creation and destruction of 
particle-antiparticle pairs of the auxiliary field and this 
removes the degeneracy giving the usual Landau levels.
It is  easy to see that $\widehat H$ does not depend on the magnetic field 
$B$. Using the completeness of Hermite functions it can be rewritten as
$m\sum_{p}( A^\dag_p A_p + B^\dag_p B_p )$. 
A decomposition similar to \eqref{mamma} holds for the free
Hamiltonian of the massive field in NJL models  without
magnetic field: the creation and destruction operators of the
massless field play the role of the hatted operators  and the mass
replaces the magnetic field in the quadratic interaction.
\subsection{\bf Magnetic pairing and magnetic catalysis}
Let us summarize what emerges from the previous analysis.
A characteristic feature of SSB with a mechanism akin to
supeconductivity is a pairing of particles and antiparticles.
This is what happens in the NJL model \cite{NJL} where 
the pairing is due to the attractive nonlinear interaction
provided it is sufficiently strong. In the previous subsection
we have shown that pairing is induced in a free Dirac field in
$2+1$ dimensions by a constant magnetic field for any value
of the mass. There is no surprise therefore that when we switch on the nonlinear
interaction in the limit of zero mass chiral SSB takes place
even for very small values of the coupling. In $2+1$ dimensions
the dependence of the dynamical mass on the nonlinear coupling
is analytic as shown in  \cite{GMS}.
The point of view discussed in this paper complements the analysis of
Gusynin, Miransky and Shovkovy. The magnetic catalysis appears as a
special consequence of the more general phenomenon represented by magnetic 
pairing.

\section{Spontaneous symmetry breaking in nonequilibrium}
SSB has been studied so far mainly as an equilibrium phenomenon
typical of systems with infinitely many degrees of freedom.
It was discovered however some time ago
\cite{NEQ} that out of equilibrium SSB can take place through
mechanisms  not available in equilibrium. For example, it is
well known that in one-dimensional systems with short range interactions   
like a one-dimensional Ising model, SSB is not possible. 
However  it was shown that in one-dimensional models through which 
currents are flowing due to unequal
boundary conditions at the two sides of the system, stationary
states breaking a symmetry are possible. In the cases considered
the broken symmetry is $CP$, the product of charge conjugation
and parity. In spite of the fact that fifteen years have elapsed
the study of these models is still at the beginning and there are
very few rigorous results. This may be due to the great
difficulties encountered in non equilibrium physics or also to
the lack of other major motivations. 
One motivation of the models  apparently was provided by traffic problems,
therefore not connected with condensed matter or particle physics. 
They are very special but I wish to  emphasize 
the important message they convey: 

\medskip

{\sl Stationary states are the obvious generalization
of equilibrium states but the conditions under which SSB takes place
in nonequilibrium are different from equilibrium. 
In stationary nonequilibrium states SSB may be possible even when it is not 
permitted in equilibrium.} 

\medskip

We describe briefly the class of one-dimensional 
models introduced in \citen{PEM} which belongs to
the category of boundary driven particle systems. One considers a
one dimensional lattice of length $N$, called the bridge, where 
each point can be
occupied either by a positive $(+)$ or a negative $(-)$ particle 
or be empty $(0)$. The positive particles move to the right while
the negative particles move to the left with hard core exclusion.
At each end of the bridge there are junctions where the lattice splits
into two parallel segments, access and exit lanes, one containing only 
plus particles and empty sites
and the other only minus particles and empty sites.  

Plus particles are injected in the access lane at the left with rate
$\alpha$, if the first site is empty, and removed with rate $\beta$
from the right end of the exit lane (on the right of the lattice), if the
last site is occupied. Likewise minus particles are injected into their
access lane on the right with rate $\alpha$ and removed with rate $\beta$
from the left end of their exit lane. Plus and minus particles
hop with rate $1$ to the right and to the left, respectively, in
their access-exit lanes. Plus particles enter the bridge at the
left junction if the first site is empty and leave it at the other junction
if the first site of the exit lane is empty, with rate $1$. Inside the 
bridge plus particles exchange with empty sites and with minus
particles with rate $1$. The same rules hold for the minus particles
after exchanging right with left. The model is clearly symmetric 
with respect to simultaneous charge conjugation and left-right reflection.

\begin{figure}   
  \begin{center}
  \includegraphics[width=12cm,clip]{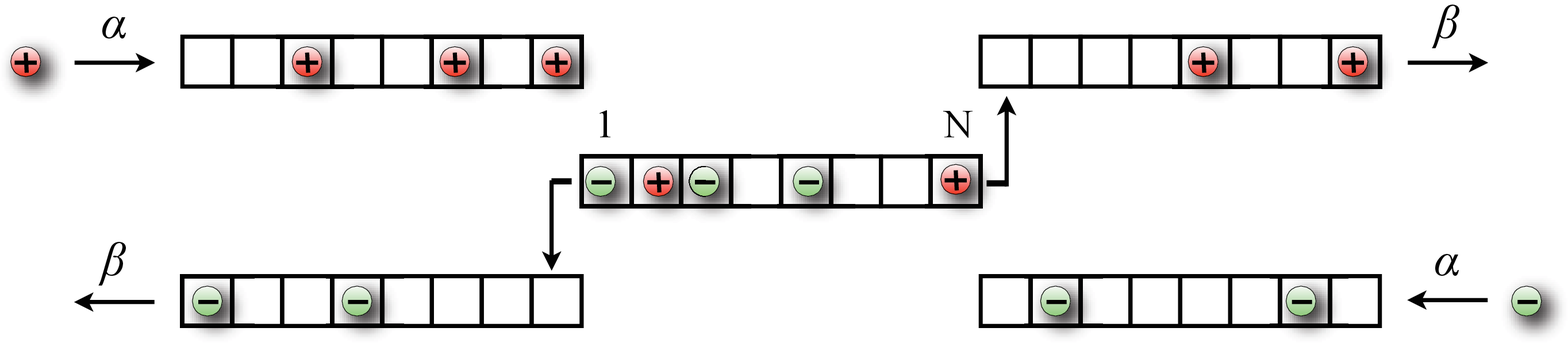}
  \end{center}
  \caption{The bridge model with access and exit lanes after \citen{PEM}}
\end{figure}

Summarizing the dynamics, during 
a time interval $dt$ three types of exchange events can take place between
two adjacent sites 
\begin{equation}
+0\rightarrow 0+ \;, \quad 0-\rightarrow -0\; , \quad +-\rightarrow -+ \;,
\end{equation}
with probability $dt$. The last one takes place only on the bridge. 
At the left of the access lane of plus particles 
we have
\begin{equation}
0\rightarrow + \;,
\end{equation} 
with probability $\alpha dt$.
At the right end of the exit lane of plus particles 
\begin{equation}
+\rightarrow 0 \;,
\end{equation}
with probability $\beta dt$,
and similarly for minus particles after reflection.
One is interested in the steady states in the thermodynamical
limit $N\rightarrow \infty$. These are characterized by 
the density profiles of the two charges and the corresponding 
currents. 

The phase diagram of this system in the $(\alpha, \beta)$ plane
exhibits three phases of which two are symmetric, i.e. the densities
and currents of the two charges are the same, and one is non symmetric.
The symmetry breaking phase exists for low values of the 
rate $\beta$. In this phase the densities and currents
of the positive and negative particles are different.
There are two stationary states  breaking the  $CP$ symmetry
in which either the positive particles or the negative particles
dominate.These states are obtained one from the other by applying
the $CP$  operation. For finite $N$ jumps between these states are
possible but one can prove that the mean jump time diverges
exponentially with $N$ so that they become stable in the
thermodynamic limit.  
For details the reader is referred to \citen{PEM}.

\subsection{\bf Possible relevance of nonequilibrium SSB}
There are  facts in the world around us that so far have eluded
a really satisfactory explanation. At the planetary scale
we know that in living matter left-handed chiral molecules
are the rule. 
The question was raised by Pasteur
in the XIXth century \cite{P}. Explanations have been attempted for
example invoking a small initial unbalance between left-handed and
right-handed molecules in the prebiotic world subsequently amplified
through various mechanisms \cite{BLA}. 
The effects of parity violation in weak interations also have been
considered \cite{SA}. 
I think that nonequilibrium SSB in stationary states offers
another direction to be explored. Kinetic models based on 
nonequilibrium stationary states, therefore closer to the point of
view advocated here, have been proposed, see e.g. \citen{PBC}.
What is missing however is a full statistical mechanics approach to the
problem allowing to disentangle the general mechanism from
the chemical details of each model. 

\medskip

At the cosmic scale we do not understand completely why matter is so much
more abundant than antimatter. Also in this case explanations have been proposed
based on evolutionary nonequilibrium invoking small symmetry violations
which are amplified to reach the present state
of the universe. For recent reviews see \citen{YO,SH}. 
These explanations, in which
nonequilibrium is a fundamental ingredient, require a detailed
reconstruction of the histories which are complicated and with many
uncertainties. 
What I am suggesting here is another way in which nonequilibrium may 
play a role. Stationary or quasi-stationary nonequilibrium states in 
which symmetries
are spontaneously broken may appear during an evolutionary process.
The dynamics of currents is the additional ingredient which  enriches 
the picture with respect to equilibrium. This may require some departure
from the prevailing picture of the early history of our universe.

\medskip

Nonequilibrium stationary states with SSB deserve further study
both as a purely theoretical problem and in view of applications
to natural phenomena. A first step is to study this phenomenon
in dimension greater than one. 
We point out that generically
new types of phase transitions arise in nonequilibrium. 
They are 
connected with singularities of the free energy not permitted in
equilibrium \cite{BDGJL}. Moreover the thermodynamics of current fluctuations
shows that dynamical phase transitions
spontaneously breaking translation invariance in time, are possible 
\cite{BDGJL1,BD}. One should profit of this wider perspective
in the study of old and new problems.

\section*{Acknowledgements}
This text is based on a talk given at the Yukawa Institute
in Kyoto on the occasion of a meeting in honor of Yoichiro Nambu
in the Fall 2009.
I wish to express my gratitude to Professor Eguchi for the invitation
and to Professor Hayakawa  who first suggested to write this paper.
I thank Carlo Presilla for past and present collaboration 
on the first topic discussed  in this paper and for a
critical reading of the manuscript. I am grateful to Francesca Marchetti for
our past collaboration which led to the study of the second topic considered
in this paper.

\end{document}